\documentclass[12pt]{iopart}
\usepackage{iopams}
\usepackage{amsmath}
\usepackage{amssymb}
\usepackage{amsfonts}
\usepackage{verbatim}
\usepackage [margin=1in]{geometry}
\usepackage{tikz}
\usetikzlibrary{arrows}
\usepackage[multiple]{footmisc}
\usepackage{relsize}
\usepackage{bbold}
\usepackage{enumerate}
\usepackage{float}

\newcommand {\eq}[1] {(\ref{#1})}

\newcommand{\re}[1]{\mathrm{Re}(#1)}
\newcommand{\im}[1]{\mathrm{Im}(#1)}

\newcommand {\lin}[2] {\mathsmaller{\left\lfloor \frac{#1}{#2} \right\rfloor}}
\newcommand {\uin}[2] {\mathsmaller{\left\lceil \frac{#1}{#2} \right\rceil}}

\newcommand {\plus} {\left\{+\right\}}

\input epsf.sty

\newlength{\TZ}
\begin{document}
\title[Spirals and coarsening patterns in the competition of many species]{Spirals and coarsening patterns in the competition of 
many species: A complex Ginzburg-Landau approach}

\author{Shahir Mowlaei, Ahmed Roman, and Michel Pleimling}
\address{Department of Physics, Virginia Polytechnic Institute and State University, Blacksburg, Virginia 24061-0435, USA}

\begin{abstract}
In order to model real ecological systems one has to consider many species that interact in complex ways. However, most
of the recent theoretical studies have been restricted to few species systems with rather trivial interactions. The few studies
dealing with larger number of species and/or more complex interaction schemes are mostly restricted to numerical explorations.
In this paper we determine, starting from the deterministic mean-field rate equations,
for large classes of systems the space of coexistence fixed points at which biodiversity is
maximal. For systems with a single coexistence fixed point we derive complex Ginzburg-Landau equations that allow to describe
space-time pattern realized in two space dimensions. For selected cases we compare the theoretical
predictions with the pattern observed in numerical simulations.
\end{abstract}

\section{Introduction}
Understanding the generic conditions for biodiversity and species extinction remains a challenging problem in
evolutionary and population dynamics \cite{May74,Smith74,Sole06}. 
Whereas real world ecological systems are composed of tens or hundreds of species, 
theoretically well understood cases remain restricted to systems with only very few species that interact in rather simple
ways. But even these very simple cases have revealed a very rich behavior, due to the nonlinearity inherent to this type
of systems. Further progress in this field can be expected through the use of well established methods from
nonlinear dynamics and statistical physics \cite{Sza07,Fre09}.

The simplest few species models, like the three-species rock-paper-scissors
model or its four species variant, have been the subject of a range of in-depth studies that have unveiled
many generic properties of systems with cyclic competition \cite{Fra96a,Fra96b,Fra98,Kob97,Pro99,Tse01,Sat02,
Ker02,Sza04,Kir04,He05,Rei06,Sza07b,Rei07,Rei07a,Sza08,Cla08,Pel08,Rei08a,Rei08b,Ber09,Ven10,Cas10,Shi10,
And10,Wan10,Mob10,He10,Win10,Nob11,Dur11,He11,Rul11,Wan11,Nah11,Jia11,Pla11,Dem11,Zia11,He12,Don12,Dob12,
Juu12,Lam12,Jia12,Ada12,Juu12a,Dur12,Rom12,Rul13,Szs13,Int13,Gui13,Par13,Sch13}.
Similar studies 
of more complicated systems composed of multiple species interacting in less trivial ways have been scarce until recently
\cite{Sza01,Sza01b,Sza05,Sza07c,Per07,Sza08a,Van12,Lut12,Ave12a,Ave12b,Rom13,Kne13,Ave13,Vuk13,Kan13,Ave13b,Dob14}.
These few investigations of the more complicated cases have been largely restricted to the numerical exploration
of the most prominent features. However, in order to develop a better understanding of these cases, analytical approaches
are needed. As a first step in that direction we present in this paper 
some analytical results for a large class of systems that display complicated
interaction schemes. We thereby discuss different dynamics: (1) Lotka-Volterra dynamics where the number of individuals is conserved,
(2) May-Leonard dynamics where this number is no longer constant, as well as (3) a mixture of both Lotka-Volterra and
May-Leonard dynamics \cite{Rul13}.
For very general cases we investigate the deterministic mean-field rate equations and determine the space
of coexistence fixed points on which species coexist and therefore biodiversity prevails.
For cases with a single coexistence fixed point we determine the invariant manifold and study
the dynamics around this point. For two-dimensional lattices this is done through a complex Ginzburg-Landau approach that
allows to derive expressions for various quantities of interest.

Our paper is organized in the following way. After having introduced our model in section 2, we determine in section 3
the space of coexistence fixed points for very general cases. We thereby obtain that the dimensionality of that space
depends on the chosen dynamics. In section 4 we discuss in more detail cases with a single coexistence fixed point and
derive for two-dimensional systems the complex Ginzburg-Landau equations that allow to describe the dynamics in close 
vicinity to that fixed point. In order to do so we allow the particles to be mobile and diffuse on the lattice.
These results are then applied to some selected cases.
Finally, in section 5 we discuss some consequences of our results and conclude.
Some more technical aspects are discussed in the appendices.

\section{Model}

\begin{figure}[h]
\centerline{\epsfxsize=3.75in\ \epsfbox{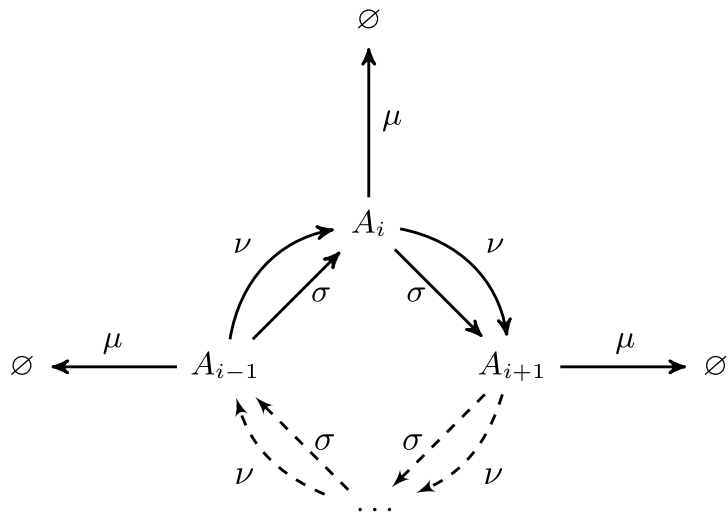}}
\caption{Schematic illustration of the $(N,1,1)$ reaction scheme.}
\label{fig1}
\end{figure}

We consider systems composed of $N$ species living on a lattice where every lattice site is either
occupied by a single individual or is empty. Individuals on neighbouring sites 
interact in the following way:
\begin{align}
A_i + \varnothing \overset{\mu}{\longrightarrow}&\;  A_i + A_i \label{eq:reaction1} \\
A_i + A_{i+j} \overset{\sigma}{\longrightarrow}&\; A_i + \varnothing \quad ; \quad 1\le j \le r \label{eq:reaction2}\\
A_i + A_{i+j} \overset{\nu}{\longrightarrow}&\; A_i + A_i \quad ; \quad 1\le j \le r' \label{eq:reaction3}
\end{align}
where $A_i$ is one individual from species $i$ and $\varnothing$ is an empty site. 
In addition we allow particles to be mobile and diffuse on the lattice by hopping to empty neighbouring sites.

In this work we only consider the case of species independent
reaction rates. Whereas the first reaction describes the birth of an off-spring with rate $\mu$, the other two reactions
describe predation events. Reaction (\ref{eq:reaction2}) is a May-Leonard type reaction where a predator simply removes
a prey from the system, thereby changing the total number of individuals. Allowing for the possibility of empty sites yields the formation
of spiral patterns, see \cite{Fre09} for a discussion of this point. Every species is thereby preying on $r$ other species
in a cyclic way, i.e. species $i$ is preying on species $i+1$, $i+2$, $\cdots$, $i+r$ (modulo $N$). The second type of 
predation (\ref{eq:reaction3}) is of Lotka-Volterra type and keeps the total number of individuals constant as a prey is 
immediately replaced by a predator. This predation happens again in a cyclic way with each species preying on $r'$ other
species. For $r'=0$ respectively $r=0$ we have a system with May-Leonard respectively Lotka-Volterra dynamics.
For the general case, where both $r$ and $r'$ are non zero, we allow for the presence of both types of dynamics.
We call this model the $(N,r,r')$ model (see figure \ref{fig1} for a
schematic illustration of the case $(N,1,1)$). The $(N,r)$ model discussed in \cite{Rom13} corresponds to the
$(N,r,0)$ model in this notation. 

\begin{figure}[h]
\centerline{\epsfxsize=3.75in\ \epsfbox{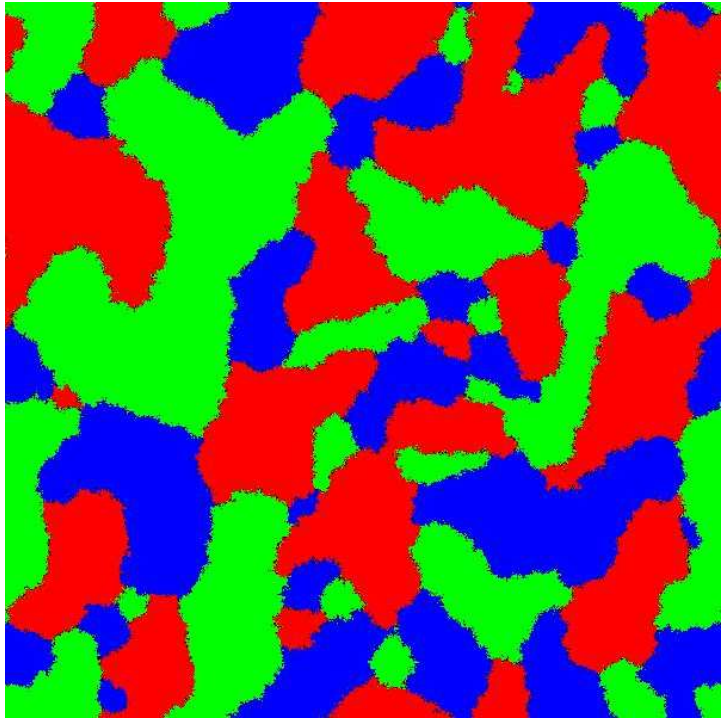}}
\caption{Coarsening pattern emerging in the (3,2,0) model in absence of mobility where the following rates 
have been used, see equations (\ref{eq:reaction1}) - (\ref{eq:reaction2}): $\sigma = 0.9$, $\mu =1$, and
$\nu = 0$. The lattice used here and in the following figures has 
$600 \times 600$ sites.}
\label{fig2}
\end{figure}

\begin{figure}[h]
\centerline{\epsfxsize=3.75in\ \epsfbox{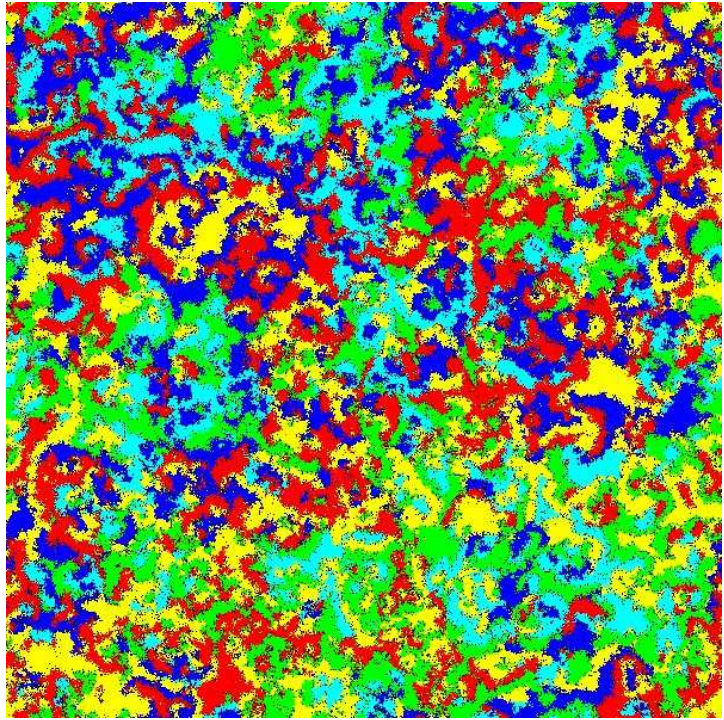}}
\caption{Complex pattern resulting from the (5,2,0) model in absence of mobility. The rates were
$\sigma = 0.9$, $\mu =1$, and $\nu = 0$.}
\label{fig3}
\end{figure}

In a spatial setting this rather simple looking interaction scheme yields a plethora of different space-time patterns. 
Figures \ref{fig2} and \ref{fig3} show two typical examples in two space dimensions (see \cite{Rom13} for other examples)
in absence of mobility.
The (3,2,0) scheme in figure \ref{fig2} provides an example of coarsening of pure domains: as every species attacks every
other species, each individual wants to be surrounded by individuals of the same species, such yielding the situation
of complete segregation. 
The (5,2,0) scheme shown in figure \ref{fig3} yields a more complex space-time pattern which results from two
different types of spirals that take place in the system at the same time. These spirals are not permanent and
break up easily. We come back to these different cases later in the paper.
We also note that other types of space time pattern can be realized (coarsening pattern where each domain contains multiple 
mutually neutral species that ally in order to fend off other alliances or coarsening pattern where inside the domains non-trivial
dynamics emerges due to the fact that the allying species are in a predator-prey relationship) and refer the interested reader to the
paper \cite{Rom13} for a discussion of these cases. All these coexistence states are quasi-stationary states: any finite 
system will eventually end up in an absorbing state where the time needed to enter this final state diverges 
with the system size \cite{Rei07a,Int13}.

In the following we first neglect any spatial degrees of freedom. We will consider the spatial dependence later 
when discussing the complex Ginzburg-Landau equations.

Introducing the time-dependent population densities $a_i$, the mean-field rate equations for the above 
reaction scheme read:
\begin{align}\label{adot}
\frac{d a_i}{dt} = a_i \left[ \mu \left(1-\sum_{j=1}^{N} a_j\right) - \sum_{j=1}^{N} \tilde{\sigma}_j a_{(i-j)} -  
\sum_{j=1}^{N} \tilde{\nu}_j \left( a_{(i-j)} - a_{(i+j)} \right) \right]~,
\end{align}
where the first term describes the birth of off-springs in presence of empty sites, whereas the other terms
result from the predation events. 
The index $(i)$ means 
\begin{align}\label{parentheses}
(i) = \left\{ \begin{array}{ll}
i ~\mbox{mod}~N & \mbox{if}~i~\mbox{is not a multiple of}~N \\
N & \mbox{otherwise}
\end{array} \right.
\end{align}
and takes care of the cyclic nature of our reaction scheme (we use 
this notation for indices throughout the paper). In addition, we have also introduced 
the shorthand notations $\tilde{\sigma}_j \equiv \sigma\;  \theta\left[r-j\right]$ and $\tilde{\nu}_j \equiv \nu\; \theta\left[r'-j\right]$, where
$\theta$ is the discrete Heaviside step function, thus indicating the possible preys for each species.

Using the identities
\begin{align}
\sum_{j=1}^{N} \tilde{\sigma}_j a_{(i-j)} & = \sum_{j=1}^{N} \tilde{\sigma}_{(i-j)} a_j \\
\sum_{j=1}^{N} \tilde{\nu}_j a_{(i+j)} & = \sum_{j=1}^{N} \tilde{\nu}_{(j-i)} a_j
\end{align}
and introducing the population density vector $\vec{a} = \left( a_1, \cdots a_N \right)^{\, T}$ allows us to cast the rate equations in the compact
form
\begin{align}\label{adotvec}
\frac{d a_i}{dt} = \mu a_i - a_i \, \left( {\mathbf F} \, \vec{a} \right)_i~,
\end{align}
where the elements of the matrix ${\mathbf F}$ are given by
\begin{align}\label{fij}
F_{ij} \equiv \mu + \tilde{\sigma}_{(i-j)} +\tilde{\nu}_{(i-j)} - \tilde{\nu}_{(j-i)}~.
\end{align}
Equations (\ref{adotvec}) and (\ref{fij}) form the starting point for the following discussion.

\section{The space of coexistence fixed points}
In this section we discuss the set of coexistence fixed points as a function of
the total number of species $N$ and of the number of preys $r$ and $r'$ each species has.
We denote as coexistence fixed point those fixed points for which none of the 
species is extinct. The reader should note that for this section it is not needed to have mobile particles.
Diffusing particles will be used when we discuss in section 4.3 the complex Ginzburg-Landau equation for
spatial systems.

In general, as we allow for empty sites, our system evolves within a $N+1$ simplex where each vertex represents the case where
one species (or empty sites) completely fills the system. These vertices are absorbing points from which the system described by
the equations (\ref{adotvec}) and (\ref{fij}) can not escape. Needless to say, that these absorbing points represent the
complete loss of biodiversity. 

The coexistence fixed points, on the other hand, are the steady states where biodiversity is maximal as all species remain
present in the system. Setting the left hand side of equation (\ref{adotvec}) to zero, these points are given by the
solutions of the equation ($\vec{\mu}$ being the vector which has the birth rate $\mu$ as each element)
\begin{align} \label{eqF}
{\mathbf F} \, \vec{a}^{\, *} = \vec{\mu}~,
\end{align}
where the vector $\vec{a}^{\, *}$ has only non-zero elements. 

As the matrix ${\mathbf F}$ is circulant, its (unnormalized) eigenvectors can be given as
\begin{align}
\vec{\Omega}_k = (\omega^{-k}, \omega^{-2k}, \cdots, \omega^{-(N-1)k}, \omega^{-Nk})^T ~~, ~~k = 1, \cdots,N
\end{align}
with $\omega = e^{\frac{2 \pi i}{N}}$. In addition, its eigenvalues are 
\begin{align}
\Lambda_k = \sum\limits_{l=1}^N  F_{1,l} \, \omega^{-(l-1)k} = \sum_{j=1}^{N}   F_{1,N-j+1} \, \omega^{jk} ~~, ~~k = 1, \cdots,N~.
\end{align}
After some algebraic manipulations (see Appendix A), the last equation can be written in the following form:
\begin{align}\label{lambda-k}
\Lambda_k 
 =&\; \left( N \mu + r \sigma \right) \delta_{k,N} + \left\{ \frac{\omega^k}{1-\omega^k} \left[ \left( 1-\omega^{rk} \right) \sigma + 
\left(1- \omega^{-(r'+1)k} \right) \left( 1- \omega^{r'k} \right) \nu \right] \right\} (1-\delta_{k,N})
\end{align}
where $\delta_{k,N}$ is the Kronecker delta.

Inspecting the expression (\ref{lambda-k}) for the eigenvalues, one sees that (a) the eigenvalue $\Lambda_N$ is always
larger than zero and that (b) the eigenvalue $\Lambda_k$ with $k \neq N$ is zero if and only if one of the following three conditions
is fulfilled:
\begin{align}
& N ~\mbox{is a divisor of}~r k~\mbox{and of}~r'k; \label{cond1} \\
& N ~\mbox{is a divisor of}~r k~\mbox{and of}~(r'+1) k; \label{cond2} \\
& \frac{N}{2}+(r'k)+\left(-(r'+1)k\right)-(rk) = 0 ~\mbox{or}~2 N \label{cond3}
\end{align}
where in the last condition the parentheses $( \cdots )$ are to be understood in the sense of convention (\ref{parentheses}).
One immediately remarks that this last condition cannot be satisfied for an odd number of species $N$.

\subsection{The case $\nu =0$}

For the case $\nu =0$, where the dynamics is exclusively of May-Leonard type, the expression (\ref{lambda-k}) for the eigenvalues reduces to
\begin{align} \label{lambda2}
\Lambda_k =\left( N \mu + r \sigma \right) \delta_{k,N} + \frac{\omega^k}{1-\omega^k}  \left( 1-\omega^{rk} \right) \sigma
\end{align}
so that the condition for vanishing $\Lambda_k$ is given by the condition that $N$ is a divisor of $r k$ or,
equivalently, that
\begin{align} \label{eqk}
k = m \left( \frac{N}{\mbox{gcd}(N,r)} \right)~~; ~~ m = 1, 2, \cdots, \mbox{gcd}(N,r)-1~.
\end{align}
Introduction of the matrix 
\begin{align} \label{eqmatOmega}
{\mathbf \Omega} = \left( \vec{\Omega}_1, \vec{\Omega}_2 , \cdots, \vec{\Omega}_N \right) 
\end{align}
and its inverse ${\mathbf \Omega^{-1}}$ with elements $\left( {\mathbf \Omega^{-1}} \right)_{ij} = \frac{\omega^{ij}}{N}$
allows us to rewrite equation (\ref{eqF}) as
\begin{align}
\left(  {\mathbf \Omega^{-1}} \cdot {\mathbf F} \cdot  {\mathbf \Omega} \right) \cdot 
\left(  {\mathbf \Omega^{-1}} \cdot \vec{a}^{\, *} \right) = {\mathbf \Omega^{-1}} \cdot \vec{\mu} 
\end{align}
and therefore 
\begin{align} \label{lambda}
{\mathbf \Lambda} \cdot \left(  {\mathbf \Omega^{-1}} \cdot \vec{a}^{\, *} \right) = \left( 0, \cdots, 0, \mu \right)^T
\end{align}
where ${\mathbf \Lambda}$ is the diagonal matrix with the eigenvalues $\Lambda_k$ of ${\mathbf F}$.

It follows from equation (\ref{lambda}) that the dimension $d$ of the space of coexistence fixed points is determined
by the rank of ${\mathbf \Lambda}$ (we give some additional details regarding this space of coexistence fixed points
in Appendix B):
\begin{align}
d = N - \mbox{rank}\left({\mathbf \Lambda}\right)~.
\end{align}
The rank of ${\mathbf \Lambda}$ is readily obtained from the condition (\ref{eqk}) under which the eigenvalues $\Lambda_k$ vanish:
\begin{align}
 \mbox{rank}\left({\mathbf \Lambda}\right) = N - \left[ \mbox{gcd}(N,r)-1 \right]~,
\end{align}
and we finally obtain that for $\nu =0$ and $\sigma > 0$ the dimension of the space of coexistence fixed points is
\begin{align} \label{d_nu_0}
d_{\sigma} = \mbox{gcd}(N,r)-1~.
\end{align}

\subsection{The case $\sigma = 0$}
A similar analysis as for the previous case immediately yields the following result for the dimension of the space of coexistence
fixed points for the case $\sigma =0$ and $\nu > 0$ where we have only Lotka-Volterra type dynamics:
\begin{align} \label{d_sigma_0}
d_\nu = \mbox{max} \left[ \mbox{gcd}(N,r'), \mbox{gcd}(N, r'+1) \right] -1~.
\end{align}

\subsection{The general case: $\nu > 0$ and $\sigma > 0$} 
As mentioned previously, for the general case where $\nu > 0$ and $\sigma > 0$ an eigenvalue $\Lambda_k$ vanishes if anyone of
the three conditions (\ref{cond1}), (\ref{cond2}), or (\ref{cond3}) is fulfilled. The conditions (\ref{cond1}) and (\ref{cond2}) are
dealt with in a very straightforward way, following the same steps as for the previous cases where one of the rates is zero, and one
obtains that the dimension of the space of coexistence fixed points is then given by
\begin{align} \label{d_sigma_nu}
d_{\sigma\nu} = \text{max}\left\{(N,[r,r']^{\equiv N}),(N,[r,r'+1]^{\equiv N})\right\}-1
\end{align}
where $[a,b] \equiv\; \text{lcm }(a,b)$ and $[a,b]^{\equiv N} \equiv\; \theta\left[(N-1)-[a,b]\right]\cdot [a,b] + \theta\left[[a,b]-N\right]$.

The condition (\ref{cond3}), which can not be fulfilled if $N$ is odd, needs to be treated separately. Assuming $N$ even, close inspection
of (\ref{cond3}) reveals that none of the proper divisors $k$ of $N$ (which are greater than 1) can satisfy this condition. Thus, if
(\ref{cond1}) and (\ref{cond2}) are not fulfilled, the space of coexistence fixed points is zero-dimensional. This case therefore is
also captured by the expression (\ref{d_sigma_nu}).

Summarizing this part of the paper, we find that the dimensionality of space of coexistence fixed points depends on the way the predation
events are implemented. This result highlights the importance of the chosen dynamics and further illustrates that conserved and
non-conserved dynamics can yield very different results in systems that are otherwise identical.

\section{The dynamics around single coexistence fixed points}
In systems with a single coexistence fixed point interesting theoretical insights can be gained by studying the motion around this
fixed point. For that the invariant manifold, i.e. the subspace in phase space which is left invariant by the deterministic rate equations,
has to be identified. Adding then the spatial degrees of freedom neglected until now yields a system of complex Ginzburg-Landau equations
that contain information on the spreading velocity of traveling waves or the wavelength and frequency of spiral waves. In \cite{Rei08b}
this calculation has been performed for the three-species cyclic May-Leonard model which corresponds to the $(3,1,0)$ case in the notation
adopted in our paper (see \cite{Fre09} for some additional details). In the following we expand this calculations to large classes
of systems with rather general interaction schemes, provided the space of coexistence fixed points is zero-dimensional.

\subsection{The cyclic model with $r=r'=1$ and an odd number of species}
We start our discussion with the case of an odd number of species $N$ that interact in a cyclic way such that $r=r'=1$.
It then follows that $\mbox{gcd}(N,r) = \mbox{gcd}(N,r') = \mbox{gcd}(N,r'+1) =1$, so that there exists only a single coexistence fixed
point, see the previous section. The steady state equation (\ref{eqF}) readily yields the reactive fixed point
\begin{align}
\vec{a}^{\, *}=\frac{\mu}{N \mu + \sigma} (1,\ldots ,1)^T
\end{align}
where all species equally coexist.

Introducing the coordinates $x_i = a_i - a_i^*$ we can rewrite the rate equations as
\begin{align} \label{eqx}
\dot x_i = - (x_i + a^*_i)\; \sum\limits_{j=1}^N F_{ij}\; x_j = -a^*_i\; \sum\limits_{j=1}^N  F_{ij} \; x_j
- x_i \sum\limits_{j=1}^N  F_{ij} \; x_j = \sum\limits_{j=1}^N A_{ij} \; x_j - G_i
\end{align}
with the matrix
\begin{align}
{\bf A}=-\frac{\mu}{N\mu+\sigma} \left( \begin{array}{cccccc}
\mu & \mu-\nu & \mu & \cdots & \mu & \mu+\sigma+\nu \\
\mu+\sigma+\nu & \mu & \mu-\nu & \cdots & \mu & \mu\\
 \vdots & \vdots & \vdots & \ddots & \vdots & \vdots \\
\mu-\nu & \mu & \mu & \cdots & \mu+\sigma+\nu & \mu
\end{array} \right)
\end{align}
and
\begin{align}
G_i= x_i \left(\mu \left(\sum_{j=1}^{N} x_j\right) + (\sigma+\nu) x_{(i-1)} - \nu x_{(i+1)} \right)
= x_i \sum_{j=1}^{N} x_j \left( \mu + (\sigma+\nu) \delta_{j,(i-1)} -\nu \delta_{j,(i+1)} \right)~.
\end{align}

The matrix ${\bf A}$ is diagonalized by means of the matrix ${\bf \Omega}$, see (\ref{eqmatOmega}):
\begin{align}
{\bf J} \equiv {\bf \Omega}^{-1} \cdot {\bf A} \cdot {\bf \Omega} = \mathrm{diag}\; (\{ \lambda_k \})
\end{align}
with the eigenvalues 
\begin{align} \label{lambda_k}
\lambda_k =&\; \sum_{j=1}^N \omega^{jk} A_{1,N-j+1}\; = \; - a^*_k \Lambda_k \notag\\
=&\; - \, \frac{\mu}{N\mu+\sigma} \left\{ (N\mu+\sigma)\; \delta_{k,N} + ( \sigma\; \omega^k + 2i \nu\; \im{\omega^k} )\; \left( 1 -\delta_{k,N} \right) \right\}~.
\end{align}
We note that $\lambda_N = - \mu < 0$, whereas the other eigenvalues form pairs that are complex conjugate: $\lambda_k = \bar{\lambda}_{N-k}$. 

Defining the new complex coordinates $\vec{z} = {\bf \Omega}^{-1} \, \vec{x}$ allows to recast equation (\ref{eqx}) as
\begin{align} \label{dotz}
\dot{\vec{z}} = {\bf J} \, \vec{z} - \vec{H}
\end{align}
with 
\begin{align} \label{H_i}
H_i = N \mu z_i z_N + \sum\limits_{j = 1}^N \left(\sigma \omega^j + 2 i \nu \im{\omega^j} \right) z_j z_{(i-j)}
\end{align}
where we used that $\sum\limits_{k=1}^N \omega_{kj} = N \delta_{j,N}$. For the $N$-th component, this expression reduces
to
\begin{align} \label{H_N}
H_N = N (\mu+\sigma) z_N^2 + 2 \sigma \sum_{j=1}^{\frac{N-1}{2}} \re{\omega^j} |z_j|^2
\end{align}
as $z_{N-k} = \bar{z}_k$.

For the cyclic model under investigation the (unstable) invariant manifold is locally spanned by the directions of eigenvalues with positive real parts (the
{\it positive} or {\it unstable} directions),
i.e. to lowest order the invariant manifold is the plane normal to the eigendirections corresponding to eigenvalues with negative real parts
(the {\it negative} or {\it stable} directions).
Whereas for the previously studied case $N=3$ \cite{Rei08b,Fre09} as well as for the case $N=5$ the invariant manifold is two-dimensional
as only one pair of conjugate eigenvalues has a negative real part,
the situation is more complicated for larger values of $N$ where multiple eigenvalues have negative real parts.

In order to determine the invariant manifold we seek functions of the form $z_k = f(\{z_l\})$ where $k$ indicates a negative direction whereas
the set $\{z_l\}$ is the set of coordinates in the positive directions: 
\begin{align} \label{eqinvman}
\dot{z}_k= \sum_{l=\alpha(k)}^{\beta(k)} \left( \dot{z}_l \partial_{z_l} \right) z_k~.
\end{align}
The summation boundaries are discussed in Appendix C.
The full problem is obviously of a formidable nature. Luckily, we only need
expressions up to second order in the $z_l$'s. We therefore make the ansatz
\begin{align} \label{eqman}
z_k = \sum_{l=\alpha(k)}^{\beta(k)} \gamma_l^k z_l z_{(k-l)}~.
\end{align}
Inserting this ansatz together with the expressions (\ref{dotz}) and (\ref{H_i}) into the equation (\ref{eqinvman}) for the invariant 
manifold yields
\begin{align}
\sum_{l=\alpha(k)}^{\beta(k)} \gamma_l^k \left( \lambda_l + \lambda_{(k-l)} \right) z_l z_{(k-l)}
= \lambda_k \sum_{l=\alpha(k)}^{\beta(k)} \gamma_l^k z_l z_{(k-l)} -  \sum_{l=\alpha(k)}^{\beta(k)} \left( \sigma\; \omega^l + 2 i \nu\; \im{\omega^l} \right) z_l z_{(k-l)}
\end{align}
from which follows that
\begin{align}
\gamma_l^k = \frac{\sigma\; \omega^l + 2 i \nu\; \im{\omega^l}}{\lambda_k - \left( \lambda_l + \lambda_{(k-l)} \right)}~.
\end{align}
Putting this back into (\ref{eqman}) leads to the equations 
\begin{align}\label{anz+}
z_k = \sum_{l=\alpha(k)}^{\beta(k)} \frac{\sigma\; \omega^l + 2 i \nu\; \im{\omega^l}}{\lambda_k - \left( \lambda_l + \lambda_{(k-l)} \right)}\; z_l z_{(k-l)}
\end{align}
that specify the invariant manifold up to quadratic order.

\subsection{The general case}
In general, whenever the space of coexistence fixed point is zero-dimensional we have that $\lambda_k = -a^*_k \Lambda_k$, since by symmetry all
components of the coexistence fixed point are identical:
\begin{align}
\vec{a}^{\, *}=\frac{\mu}{N \mu + r \sigma} (1,\ldots ,1)~.
\end{align}
Therefore for the general case $(N,r,r')$ with a single coexistence fixed point the eigenvalue $\lambda_k$ takes the form
\begin{align} \label{lambdak}
\lambda_k = -\mu \delta_{k,N} - \frac{\mu}{N \mu + r \sigma} \left\{ \frac{\omega^k}{1-\omega^k} \left[ \left( 1-\omega^{rk} \right) \sigma + 
\left(1- \omega^{-(r'+1)k} \right) \left( 1- \omega^{r'k} \right) \nu \right] \right\} \left( 1 - \delta_{k,N} \right)
\end{align}
Following the steps in the previous subsection, one can then write down for the $H_i$ 
expressions similar to those obtained for the case with $r=r'=1$ (see equations (\ref{H_i}) and (\ref{H_N})):
\begin{align}
H_i =&\; N \mu z_i z_N + \sum_{v=1}^N \left( \sigma \sum_{m=1}^r \omega^{mv} + 2i \nu \sum_{m'=1}^{r'} \im{\omega^{m'v}} \right) z_v z_{(i-v)} \label{H^r}\\
H_N =&\; N \mu z_N^2 + 2 \sigma \sum_{v=1}^{\frac{N-1}{2}} \left( \sum_{m=1}^r \re{\omega^{mv}} \right) |z_v|^2 \label{H^N}
\end{align}
More problematic are the steps that deal with the invariant manifold and the dynamics on this manifold. Indeed, although the $\lambda_k$'s
are all non-zero, their real part might be zero depending on the combination of the parameters $N$, $r$, $r'$, and $k$ as well as 
on the values of the rates $\sigma$ and $\nu$. In such cases one needs to perform a further center manifold reduction and go beyond the 
linear order for the determination of the converging or diverging behavior of those specific directions near the coexistence fixed point.

From now on we exclusively focus on those cases for which all $\lambda_k$'s have a nonvanishing real part.
Under this additional constraint the local specification of the invariant manifold is straightforward. As before 
we project the dynamics onto the unstable manifold by expressing the negative or stable directions (i.e. $k$'s with $\re{\lambda_k}<0$), 
up to quadratic order, in terms of the positive or unstable directions. This leads to the expression
\begin{align} \label{z_k_gen}
z_k =&\; \sum_{l,(k-l) \in \plus} \gamma^k_l\; z_l z_{(k-l)} 
\end{align}
where the shorthand notation
\begin{align}
\plus \equiv&\; \left\{ k\;\in\{1,2,\cdots,N\}\; : \; \re{\lambda_k}>0 \right\}
\end{align}
means the set of unstable (positive) directions, i.e. the sum is over terms where both $l$ and $(k-l)$ label unstable
directions. In equation (\ref{z_k_gen}) the parameters $\gamma^k_l$ are given by
\begin{align}
\gamma^k_l \equiv&\; \frac{f(l)}{\lambda_k - \left(\lambda_l + \lambda_{(k-l)} \right)}
\end{align}
with
\begin{align}
f(l) \equiv&\; \sigma \sum_{m=1}^r \omega^{ml} + 2i \nu \sum_{m'=1}^{r'} \im{\omega^{m'l}}~.
\end{align}

In vicinity of the unstable fixed point, this brings the rate equations on the unstable manifold to the following form ($s\in\plus$)
\begin{align}\label{rate-uns}
\dot{z}_s =&\; \lambda_s z_s - H_s  \notag\\
=&\; \lambda_s z_s - \left\{ N \mu z_s z_N + \sum_l f(l)\; z_l z_{(s-l)} \right\}~.
\end{align}

The next step is to suppress the quadratic terms in the above equations. This is achieved by means of a family of near identity transformations as
\begin{align}
z_s \mapsto z_s + h_2^s(\{z\})
\end{align}
where $h_2^s(\{z\})$ are quadratic polynomials in the coordinates of the unstable directions $s$ \cite{wiggins}:
\begin{align}\label{hr2}
h_2^s(\{z\}) = \sum_{u,t \;\in\plus} a^s_{ut} z_u z_t~.
\end{align}
This transformation will dismiss the quadratic terms (with suitable choice of $a^s_{ut}$) at the price of introducing further cubic terms. 
Actually, a subset of these cubic terms are welcome as we aim at ending up with a family of Stuart-Landau normal forms. 
The rest, however, is redundant and can again be suppressed by another family of near identity transformations
\begin{align}\label{hr3}
z_s \mapsto z_s + \bar{h}^s_3(\{z\}) 
\end{align}
where $\bar{h}^s_3(\{z\})$ is cubic in coordinates corresponding to the unstable directions: 
\begin{align}
\bar{h}^s_3(\{z\}) \equiv \sum_{v,t,u \;\in\plus} \bar{a}^s_{vtu} z_v z_t z_u~.
\end{align}
Here the bar indicates that the $z_s |z_s|^2$ terms, 
needed for the Stuart-Landau normal forms, are preserved in the process of transforming the rate equations (\ref{rate-uns}).

Careful performance of transformations \eq{hr2} and \eq{hr3} then results in the following set of Stuart-Landau normal forms 
\emph{on the unstable manifold} ($s\in\plus$):
\begin{align}\label{zs3+}
\dot{z}_s =&\, \lambda_s z_s - \left\{ \frac{2\;\re{f(s)} \cdot \left\{ \left(N\mu+r\sigma\right) +
f(s) \right\}}{\lambda_N-2\;\re{\lambda_s}} \; + \; \frac{f(s)\cdot\left\{f(2s)+f(-s)\right\}}{\lambda_{(2s)} - 2 \lambda_s} \right\} z_s |z_s|^2
\nonumber \\
=&\, \lambda_s z_s - G(s) z_s |z_s|^2~.
\end{align}
Restricting ourselves to the directions with negative imaginary part of the eigenvalue, this can be brought in the standard form
\begin{align}\label{zs3}
\dot{z}_s = (c_{1,s} - i \omega_s) z_s - c_{2,s} (1+i c_{3,s})\; z_s |z_s|^2 
\end{align}
with
\begin{align}
c_{1,s} \equiv&\; \re{\lambda_s} \\
\omega_s \equiv& - \im{\lambda_s} \\
c_{2,s} \equiv&\; \re{G(s)} \\
c_{3,s} \equiv&\; \frac{\im{G(s)}}{\re{G(s)}} \label{c-3}
\end{align}

\subsection{The complex Ginzburg-Landau equations}
Recalling that we allow for mobile particles, we want to incorporate at this stage the mobility of particles that can diffuse
by jumping to an empty neighbouring site. Following \cite{Rei08b} we then obtain in the continuum limit the following reaction-diffusion
equations for the space- and time-dependent particle densities:
\begin{align}\label{rd}
\frac{d a_i(\vec{r},t)}{dt} =&\, D \nabla^2 a_i(\vec{r},t) + a_i(\vec{r},t) \left[ \mu \left(1-\sum_{j=1}^{N} a_j(\vec{r},t)\right) - \sum_{j=1}^{N} \tilde{\sigma}_j a_{(i-j)}(\vec{r},t) \right. \nonumber \\
&\, \left. - \sum_{j=1}^{N} \tilde{\nu}_j \left( a_{(i-j)}(\vec{r},t) - a_{(i+j)}(\vec{r},t) \right) \right]~,
\end{align}
that differ from the rate equations (\ref{adot}) by the diffusion term $D \nabla^2 a_i(\vec{r},t)$, where $D$ is the diffusion constant that depends on the
rate for jumps into unoccupied neighbouring sites. Note that the realization of mobility through jumps into empty sites
yields nonlinear diffusive terms in addition to the usual linear term \cite{Szs13}. 
As we expect the dynamics to be dominated
by the long wavelength modes, we only keep the leading order gradient term in equation (\ref{rd}).

When redoing the calculation of the previous subsection with this equation, one notes the appearance of nonlinear diffusive terms due
to the nonlinearities of the coordinate transformations. These nonlinear terms are expected to be subleading
when the dynamics is dominated by the long wavelength modes \cite{Rul13}. 
We therefore ignore these additional terms and end up with the following system of 
partial differential equations:
\begin{align}\label{diff}
\dot{z}_s = D\; \nabla^2 z_s + (c_{1,s} - i \omega_s)\; z_r - c_{2,s} (1+i c_{3,s})\; z_s |z_s|^2 ~,
\end{align}
where $s$ labels the unstable directions with negative imaginary parts.
This is a set of complex Ginzburg-Landau equations similar to those that have been studied extensively in the past
in a large variety of different physical situations \cite{Cro93,Ara02}. It should be stressed that neglecting the
nonlinear diffusive terms is an uncontrolled approximation and that {\it a posteriori} tests are needed to
check how reliable this approximation really is. For the three-species case (3,1,0), where the same assumptions
have been made, it was checked through numerical simulations that the resulting complex Ginzburg-Landau equation
faithfully describes some of the main features of the corresponding lattice model \cite{Rei07,Rei07a,Rei08b}.
Similar tests for checking the reliability of this approximation should also be done for the larger class
of models studied in our paper, and we plan to come back to this important aspect in the future.

From these equations a variety of quantities can be computed. First we note that in the case $c_{3,s} = 0$ and $\omega_s =0$ the complex Ginzburg-Landau 
equation reduces to the time-dependent real Ginzburg-Landau equation used commonly for the description of coarsening in ferromagnets \cite{Bra02}
(see \cite{Pat13} for a comparison between the dynamics in the complex and the real Ginzburg-Landau equations). For the case $c_{3,s} \ne 0$ and
$\omega_s \ne 0$ the corresponding complex Ginzburg-Landau equation allows to describe spiral waves and derive analytical expressions
for a range of relevant quantities \cite{Cro93,Ara02,Fre09}. For example for the linear spreading velocity $v^*_s$ one obtains
\begin{align}
v^*_s =&\; 2 \sqrt{c_{1,s} D}~,
\end{align}
whereas the wavelength $\lambda^*_s$ and the frequence $\Omega_s$ of the spiral are given by
\begin{align}
\lambda^*_s =&\; \frac{2\pi c_{3,s} \sqrt{D}}{\sqrt{c_{1,s}} \left( 1- \sqrt{1+c^2_{3,s}} \right)} \label{lambda-star} \\
\Omega_s =&\; \omega_s + \frac{2 \pi v_s^*}{\lambda_s^*}
\end{align}

\subsection{Applications}

Let us now apply this formalism to the two examples given in section 2. 

For the case (3,2,0), see figure \ref{fig2}, we have only one unstable direction. A straightforward calculation then yields the expressions
\begin{align}
c_{1,1} = &\; \frac{\mu\, \sigma}{3\, \mu + 2\, \sigma} \\
\omega_1 =&\; 0 \\
c_{2,1} =&\; -\frac{6\, \sigma^2\, \left(3\, \mu + 2\, \sigma\right)}{\mu\, \left(3\, \mu + 4\, \sigma\right)} \\
c_{3,1}=&\; 0
\end{align}
for the four parameters in equation (\ref{diff}). As $\omega_1 = c_{3,1} = 0$ the complex equation reduces to the real one.
Consequently, one expects for this interaction scheme the appearance of coarsening domains, in complete agreement with the 
pattern showing up in figure \ref{fig2}.


The case (5,2,0), see figure \ref{fig3}, represents a more complicated situation with two unstable directions $s=1$ and $s=3$ and
two complex Ginzburg-Landau equations. 
Applying our formalism to this case yields the following expressions:
\begin{align}
c_{1,1} =&\; \frac{\mu\, \sigma}{10\, \mu + 4\, \sigma} \\
\omega_1 =&\; \frac{\mu\, \sigma\, \left(\sqrt{2} + \sqrt{10}\right)\, \sqrt{\sqrt{5} + 5}}{40\, \mu + 16\, \sigma} \\
c_{2,1} =&\; \frac{\sigma\, \left(5\, \mu + 2\, \sigma\right)\, \left( (25- 5\, \sqrt{5}\,)\, \mu + (9- 3\, \sqrt{5}\,)\, \sigma  \right)}{4\, \mu\, \left(5\, \mu + 3\, \sigma\right)} \\
c_{3,1} =&\; \frac{\left(5\, \sqrt{5} - 13\right)\, \sqrt{\sqrt{5} + 5}\, \left( (65\, \sqrt{2}\, + 25\, \sqrt{10}\,)\, \mu + (20\, \sqrt{2}\, + 6\, \sqrt{10}\,)\, \sigma\right)}{44\, \left( (25- 5\, \sqrt{5}\,)\, \mu + (9- 3\, \sqrt{5}\,)\, \sigma\right)} \\
c_{1,3} =&\; \frac{\mu\, \sigma}{10\, \mu + 4\, \sigma} \\
\omega_3 =&\; \frac{\mu\, \sigma\, \sqrt{\sqrt{5} + 5}\, \left(3\, \sqrt{2} - \sqrt{10}\right)}{40\, \mu + 16\, \sigma} \\
c_{2,3} =&\; \frac{\sigma\, \left(5\, \mu + 2\, \sigma\right)\, \left( (25+ 5\, \sqrt{5}\,)\, \mu + (9+ 3\, \sqrt{5}\,)\, \sigma\right)}{4\, \mu\, \left(5\, \mu + 3\, \sigma\right)} \\
c_{3,3} =& \frac{\left(5\, \sqrt{5} + 13\right)\, \sqrt{\sqrt{5} + 5}\, \left( (45\, \sqrt{10}\, - 95\, \sqrt{2}\,)\, \mu  + (13\, \sqrt{10}\, - 25\, \sqrt{2}\,)\, \sigma\right)}{44\, \left( (25+ 5\, \sqrt{5}\,)\, \mu + (9+ 3\, \sqrt{5}\,)\, \sigma\right)}
\end{align}
One notes that $c_{1,1}=c_{1,3}$ but $c_{3,1} \neq c_{3,3}$. This describes a situation of two different types of spirals
with the same spreading velocities but different wavelengths. Close inspections of snapshots like that shown in figure \ref{fig3}
reveals indeed the presence of two different types of spirals, characterized by different thicknesses of their spiral arms (which 
corresponds to different wavelengths), that interfere continually.

\section{Discussion and conclusion}
In this paper we have derived complex Ginzburg-Landau equations for $N$ species models in two dimensions with a large range
of different interaction schemes given by the reactions (\ref{eq:reaction1})-(\ref{eq:reaction3}). Depending on the values of the parameters,
a range of scenarios can be realised, as for example coarsening of pure domains, single spiral waves, or interacting multiple spiral waves.
Numerical simulations of the corresponding lattice gas models show a quantitative agreement with the predictions
that follow from the values of the parameters in the complex Ginzburg-Landau equations.
However, as neglecting the nonlinear diffusive terms that emerge when 
applying the nonlinear transformations is an uncontrolled approximation, more
advanced checks through the quantitative comparison of the predictions from
the complex Ginzburg-Landau equations and the results from lattice model 
simulations are needed in order to fully assess the reliability of this approximation.

It should be noted that recently an alternative approach was proposed that allows to derive in a more controlled way,
through a perturbative expansion, a set of complex Ginzburg-Landau equations for systems with cyclic competition and an
additional mutation process \cite{Szs13}.
This method has been used successfully for the three-species case (3,1,0). The same approach should also work for the
larger class of models discussed in our manuscript, and it would be interesting to compare our results 
with results obtained from this alternative approach.

\begin{figure}[h]
\centerline{\epsfxsize=3.75in\ \epsfbox{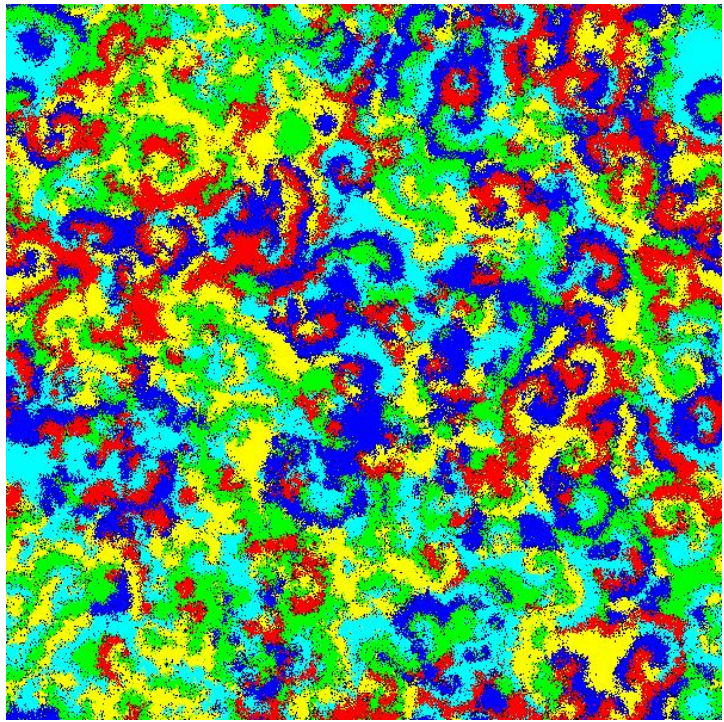}}
\caption{Stable spirals are formed in the (5,2,0) case when considering swapping. The rates were $\sigma = 0.9$, 
$\mu = 0.25$, and $\nu = 0$. In addition, particles were allowed to hop to empty neighbouring sites with rate 0.7 and
to exchange places with particles on neighbouring sites with rate 0.06.}
\label{fig4}
\end{figure}

There are two standard ways to implement mobility in systems where at most one particle is allowed at any lattice site:
diffusion, realised through the hopping of particles to empty sites, and particle swapping, where particles on neighbouring sites
exchange places. In this paper we restricted ourselves to the first case. However, the space-time pattern may change depending on the 
chosen way to implement mobility. As an example we show in figure \ref{fig4} the (5,2,0) case with swapping, which
should be compared to figure \ref{fig3} which shows a snapshot of the same system 
in absence of mobility. Using swapping yields pronounced and very stable 
spirals, characterised by wave number and frequency that differ from those obtained from the case with diffusion.
We will address the cases with swapping in a separate publication.

Obviously our numerical simulations are of stochastic nature, but we exclusively focused on deterministic equations in our theoretical
treatment. One may then wonder how the inclusion of noise, that yields a system of coupled stochastic partial differential equations,
would change our results. A first, rather trivial remark, is that in simulating
a finite system one eventually ends up with the extinction of some of the species \cite{Fre09,Int13}. We therefore focus
on earlier times that are much shorter than any extinction times. 
For the three-species May-Leonard model, the model (3,1,0) in our notation, it was found \cite{Rei08b,Fre09}
that the numerically determined spreading velocity quantitatively agrees with the prediction from the complex Ginzburg-Landau equation,
and this for a large range of values of the reaction rates. The functional dependence of the wavelength was also found to agree between
numerics and theory. However, other interaction schemes, especially for more than three species, have not been studied in the same
way. It is therefore an important open question whether a similar good quantitative agreement can be achieved for more complicated
cases. We plan to come back to this question in the future.

Another interesting question concerns the cases where at least one of the eigenvalues $\lambda_k$ (\ref{lambdak}) has a vanishing
real part. In that case an additional center manifold reduction has to be performed in order to go beyond the linear order. This is
a very demanding calculation that we leave for the future.

Although our study deals with a large class of systems, the derivation of the complex Ginzburg-Landau equations remains restricted
to cases where the space of coexistence fixed points is zero-dimensional. It is an open problem whether similar theoretical
insights can be gained for cases where that space is of higher dimension.\\[0.5cm]

\noindent
{\bf Acknowledgements:}\\~\\
This work is supported by the US National
Science Foundation through grant DMR-1205309.
We thank Uwe C. T\"{a}uber and Darka Labavi\c{c} for useful discussions.

\appendix
\setcounter{section}{0}
\section{Derivation of equation (\ref{lambda-k})}
For the explicit calculation of the eigenvalues (\ref{lambda-k}) we start from the expression
\begin{align}
\Lambda_k = \sum_{j=1}^{N}  F_{1,N-j+1} \, \omega^{jk} ~~, ~~k = 1, \cdots,N~.
\end{align}
Writing out the element of the matrix ${\mathbf F}$, we obtain
\begin{align}
F_{1,N-j+1} =&\; \mu + \tilde{\sigma}_{(1-[N-j+1])} + \tilde{\nu}_{(1-[N-j+1])} - \tilde{\nu}_{([N-j+1]-1)} \notag\\
=&\; \mu + \tilde{\sigma}_j + \tilde{\nu}_j - \tilde{\nu}_{(-j)} \notag\\
=&\; \mu + \sigma\; \theta[r-j] + \nu\; \theta[r'-j] - \nu\; \theta [r' - (N-j)]
\end{align}
and therefore
\begin{align}\label{lambda-k-}
\Lambda_k =&\; \mu \sum_{j=1}^N \omega^{jk} + \sigma \sum_{j=1}^N \omega^{jk} \theta\,[r-j] + \nu \sum_{j=1}^N \omega^{jk} \theta\,[r'-j] - \nu \sum_{j=1}^N \omega^{jk} \theta\,[r' -(N-j)] \notag\\
=&\; N \mu\, \delta_{k,N} + \sigma \sum_{j=1}^r \omega^{jk} + \nu \sum_{j=1}^{r'} \omega^{jk} - \nu \sum_{j=N-r'}^{N-1} \omega^{jk}
\end{align}
where $\delta_{k,N}$ is the Kronecker delta.
Inserting
\begin{align}
\sum_{j=1}^{r} \omega^{jk} =&\; r\, \delta_{k,N} + \omega^k \frac{1-\omega^{rk}}{1-\omega^k}\; \left(1-\delta_{k,N}\right) \label{k1} \\
\sum_{j=N-r'}^{N-1} \omega^{jk} =&\; r' \delta_{k,N} + \omega^{(N-r')k} \frac{1-\omega^{r'k}}{1-\omega^k} \left(1-\delta_{k,N}\right)\; \notag\\
=&\;  r' \delta_{k,N} + \omega^{-r'k} \frac{1-\omega^{r'k}}{1-\omega^k}\; \left(1-\delta_{k,N}\right) \label{k2}
\end{align}
into equation (\ref{lambda-k-}) finally yields the expression (\ref{lambda-k}) for the eigenvalues.

\section{Some remarks on the space of coexistence fixed points for $\nu = 0$}

Looking back at equation (\ref{lambda2}) we can see that for the $k$'s of \eq{eqk} the corresponding components 
of the vector $\mathbf{\Omega}^{-1}\cdot {\vec a}^{\, *}$ in \eq{lambda} are arbitrary. The number of these components 
is clearly $\mathrm{gcd}(N,r)-1$ and, as a result, so is the dimension of the space of coexistence fixed points. 
As for other components of this vector, the last component is simply $\mu \Lambda_N^{-1}$, and all the remaining 
components vanish since their corresponding $\Lambda_k$'s are nonvanishing in \eq{lambda}. But what does this mean 
for the vector ${\vec a}^{\, *}$ itself? To answer this question we focus on the equation
\begin{align}\label{omega-a}
\mathbf{\Omega}^{-1}\cdot {\vec a}^{\, *} = 0~.
\end{align}

One can show from the properties of the matrix $\mathbf{\Omega}^{-1}$ that, ignoring the last row $k=N$
as well as the rows $k$ such that 
$\frac{N}{D} | k$, with $D$ a proper divisor of $N$, the most general solution of the remaining rows 
in the above equation is of the form
\begin{align}
a^{\, *}_i = c_{i\, \mathrm{mod}\, D}
\end{align}
for a set of \emph{arbitrary} constants $c_j$, where $j \in \{0,1,\cdots,\frac{N}{D}-1\}$. As for the rows 
that we just ignored, this solution produces nonvanishing constants for the right hand side of \eq{omega-a} 
(so long as the constants $c_j$ are nonvanishing). But this is exactly the freedom we have for the $k$'s of \eq{eqk}, 
and therefore we can exploit this. All we need to do is to choose suitable proper divisors 
$D$ from the condition $\frac{N}{D} | k$ such that the spectrum of admissible $k$'s fits into that of \eq{eqk}. 
The possible values for $D$ are then $D_l = \frac{\mbox{{\tiny gcd}}(N,r)}{l}$ where $l$ is any natural number that satisfies $l|\mbox{gcd}(N,r)$. 
For each such $l$ then we have a family of solutions of \eq{lambda} as
\begin{align}
{\vec a}^{\, *} (l) ={\vec\alpha}^{\, *}_l
\end{align}
where
\begin{align}
\alpha^{\, *}_l : \mathbb{Z}_N \Big/ \left\langle \frac{\mbox{gcd}(N,r)}{l} \right\rangle \rightarrow 
\mathbb{R}^{\frac{\mbox{{\tiny gcd}}(N,r)}{l}}_+ \backslash G
\end{align}
The restriction to $\mathbb{R}_+$ is due to the \emph{coexistence} condition, enforcing the positivity of all 
densities. The subspace $G$ in the codomain $\mathbb{R}^{\frac{\mbox{{\tiny gcd}}(N,r)}{l}}_+ \backslash G$ is defined via
\begin{align}
G \equiv \text{ locus of } \left\{ \frac{\mbox{gcd}(N,r)}{l} \mu \Lambda^{-1}_N - \sum_{i=1}^{\frac{\mbox{{\tiny gcd}}(N,r)}{l}-1} \alpha^*_{l,i} =0 \right\}
\end{align}
and reflects the constraint on the $\alpha^{\, *}_{l,i}$'s imposed by the last row of \eq{lambda}.
$G$ is removed from $\mathbb{R}^{\frac{\mbox{{\tiny gcd}}(N,r)}{l}}_+$ to, again, ensure coexistence.

\section{The summation boundaries in the invariant manifold equations}
\begin{figure}[h]
\centerline{\epsfxsize=3.75in\ \epsfbox{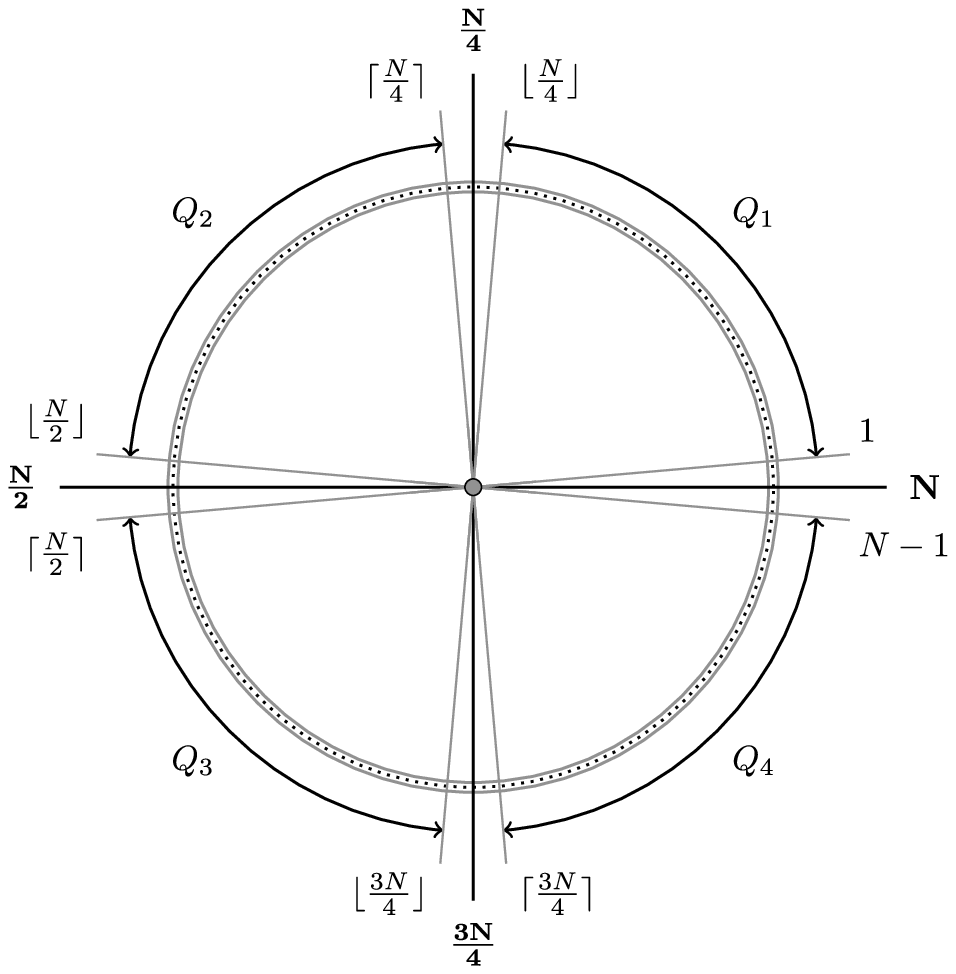}}
\caption{Schematic realization of positive and negative eigenvalues for the cyclic model with $N$ odd and $r=r'=1$.}
\label{fig5}
\end{figure}

Considering the stable directions $k$ of \eq{eqinvman} and \eq{eqman}, in order
to identify the quadratic contributions of unstable directions in the expansion $\sum_{j=1}^N (\cdots)\,z_j z_{(i-j)}$ of \eq{H_i}, 
we need to pick only those combinations $z_l z_{(k-l)}$ for which both 
$l$ and $(k-l)$ correspond to eigenvalues of \eq{lambda_k} with positive real parts. Noting that the latter equation carries an overall 
negative sign and that we are looking at the case $\nu=0$ with odd $N$, it is then straightforward to recognize the corresponding indices  
$l$ and $(k-l)$ as those belonging to the interval $\left[\uin{N}{4},\lin{3N}{4}\right]$ which is characterized as the union of quadrants 
$Q_2$ and $Q_3$ in figure \ref{fig5}.

This also implies that the set of indices that correspond to negative eigenvalues belong to the union of quadrants $Q_1$ and $Q_4$
(except for $k=N$ which exhausts the entire interval $\left[\uin{N}{4},\lin{3N}{4}\right]$). 
One can show that for $k \in Q_1$ the range of $l$ such that both $l$ and $(k-l)$ belong to $Q_2\cup Q_3$ is 
$\left[\uin{N}{4}+k, \lin{3N}{4}\right]$, whereas for $k \in Q_4$ this range is $ \left[\uin{N}{4}, \lin{3N}{4}-(N-k)\right]$. 
These two intervals can then be combined into one, namely, $\left[\alpha(k), \beta(k)\right]$ for $k\in Q_1\cup Q_4$ via the 
following identifications ($\theta$ is the discrete Heaviside step function):
\begin{align}
\alpha (k) \equiv & \uin{N}{4} + k\; \theta \left( \lin{N}{4} - k \right) \\
\beta (k) \equiv & \lin{3N}{4} - (N-k)\; \theta \left( k - \uin{3N}{4} \right)
\end{align}
Note that the special case $k=N$ is also captured by the above bounds.

\end{document}